\documentstyle[epsf,11pt]{article}
\edef\resetatcatcode{\catcode`\noexpand\@\the\catcode`\@\relax}
\ifx\miniltx\undefined\else \fi
\let\miniltx\box

\def\makeatletter{\catcode`\@11\relax}

\makeatletter

\def\@makeother#1{\catcode`#1=12\relax}

\def\@ifnextchar#1#2#3{%
  \let\reserved@d=#1%
  \def\reserved@a{#2}\def\reserved@b{#3}%
  \futurelet\@let@token\@ifnch}
\def\@ifnch{%
  \ifx\@let@token\@sptoken
    \let\reserved@c\@xifnch
  \else
    \ifx\@let@token\reserved@d
      \let\reserved@c\reserved@a
    \else
      \let\reserved@c\reserved@b
    \fi
  \fi
  \reserved@c}
\begingroup
\def\:{\global\let\@sptoken= } \:  
\def\:{\@xifnch} \expandafter\gdef\: {\futurelet\@let@token\@ifnch}
\endgroup

\def\@ifstar#1{\@ifnextchar *{\@firstoftwo{#1}}}
\long\def\@dblarg#1{\@ifnextchar[{#1}{\@xdblarg{#1}}}
\long\def\@xdblarg#1#2{#1[{#2}]{#2}}

\long\def \@gobble #1{}
\long\def \@gobbletwo #1#2{}
\long\def \@gobblefour #1#2#3#4{}
\long\def\@firstofone#1{#1}
\long\def\@firstoftwo#1#2{#1}
\long\def\@secondoftwo#1#2{#2}

\def\NeedsTeXFormat#1{\@ifnextchar[\@needsf@rmat\relax}
\def\@needsf@rmat[#1]{}
\def\ProvidesPackage#1{\@ifnextchar[%
    {\@pr@videpackage{#1}}{\@pr@videpackage#1[]}}
\def\@pr@videpackage#1[#2]{\wlog{#1: #2}}
\let\ProvidesFile\ProvidesPackage
\def\PackageInfo#1#2{\wlog{#1: #2}}

\let\DeclareOption\@gobbletwo
\let\ProcessOptions\relax

\def\RequirePackage{%
  \@fileswithoptions\@pkgextension}
\def\@fileswithoptions#1{%
  \@ifnextchar[
    {\@fileswith@ptions#1}%
    {\@fileswith@ptions#1[]}}
\def\@fileswith@ptions#1[#2]#3{%
  \@ifnextchar[
  {\@fileswith@pti@ns#1[#2]#3}%
  {\@fileswith@pti@ns#1[#2]#3[]}}

\def\@fileswith@pti@ns#1[#2]#3[#4]{%
    \def\reserved@b##1,{%
      \ifx\@nil##1\relax\else
        \ifx\relax##1\relax\else
         \noexpand\@onefilewithoptions##1[#2][#4]\noexpand\@pkgextension
        \fi
        \expandafter\reserved@b
      \fi}%
      \edef\reserved@a{\zap@space#3 \@empty}%
      \edef\reserved@a{\expandafter\reserved@b\reserved@a,\@nil,}%
  \reserved@a}

\def\zap@space#1 #2{%
  #1%
  \ifx#2\@empty\else\expandafter\zap@space\fi
  #2}

\let\@empty\empty
\def\@pkgextension{sty}

\def\@onefilewithoptions#1[#2][#3]#4{%
  \input #1.#4 }

\def\typein{%
  \let\@typein\relax
  \@testopt\@xtypein\@typein}
\def\@xtypein[#1]#2{%
  \message{#2}%
  \advance\endlinechar\@M
  \read\@inputcheck to#1%
  \advance\endlinechar-\@M
  \@typein}
\def\@namedef#1{\expandafter\def\csname #1\endcsname}
\def\@nameuse#1{\csname #1\endcsname}
\def\@cons#1#2{\begingroup\let\@elt\relax\xdef#1{#1\@elt #2}\endgroup}
\def\@car#1#2\@nil{#1}
\def\@cdr#1#2\@nil{#2}
\def\@carcube#1#2#3#4\@nil{#1#2#3}
\def\@preamblecmds{}

\def\@star@or@long#1{%
  \@ifstar
   {\let\l@ngrel@x\relax#1}%
   {\let\l@ngrel@x\long#1}}

\let\l@ngrel@x\relax
\def\newcommand{\@star@or@long\new@command}
\def\new@command#1{%
  \@testopt{\@newcommand#1}0}
\def\@newcommand#1[#2]{%
  \@ifnextchar [{\@xargdef#1[#2]}%
                {\@argdef#1[#2]}}
\long\def\@argdef#1[#2]#3{%
   \@ifdefinable #1{\@yargdef#1\@ne{#2}{#3}}}
\long\def\@xargdef#1[#2][#3]#4{%
  \@ifdefinable#1{%
     \expandafter\def\expandafter#1\expandafter{%
          \expandafter
          \@protected@testopt
          \expandafter
          #1%
          \csname\string#1\expandafter\endcsname
          {#3}}%
       \expandafter\@yargdef
          \csname\string#1\endcsname
           \tw@
           {#2}%
           {#4}}}
\def\@testopt#1#2{%
  \@ifnextchar[{#1}{#1[#2]}}
\def\@protected@testopt#1{
  \ifx\protect\@typeset@protect
    \expandafter\@testopt
  \else
    \@x@protect#1%
  \fi}
\long\def\@yargdef#1#2#3{%
  \@tempcnta#3\relax
  \advance \@tempcnta \@ne
  \let\@hash@\relax
  \edef\reserved@a{\ifx#2\tw@ [\@hash@1]\fi}%
  \@tempcntb #2%
  \@whilenum\@tempcntb <\@tempcnta
     \do{%
         \edef\reserved@a{\reserved@a\@hash@\the\@tempcntb}%
         \advance\@tempcntb \@ne}%
  \let\@hash@##%
  \l@ngrel@x\expandafter\def\expandafter#1\reserved@a}
\long\def\@reargdef#1[#2]#3{%
  \@yargdef#1\@ne{#2}{#3}}
\def\renewcommand{\@star@or@long\renew@command}
\def\renew@command#1{%
  {\escapechar\m@ne\xdef\@gtempa{{\string#1}}}%
  \expandafter\@ifundefined\@gtempa
     {\@latex@error{\string#1 undefined}\@ehc}%
     {}%
  \let\@ifdefinable\@rc@ifdefinable
  \new@command#1}
\long\def\@ifdefinable #1#2{%
      \edef\reserved@a{\expandafter\@gobble\string #1}%
     \@ifundefined\reserved@a
         {\edef\reserved@b{\expandafter\@carcube \reserved@a xxx\@nil}%
          \ifx \reserved@b\@qend \@notdefinable\else
            \ifx \reserved@a\@qrelax \@notdefinable\else
              #2%
            \fi
          \fi}%
         \@notdefinable}
\let\@@ifdefinable\@ifdefinable
\long\def\@rc@ifdefinable#1#2{%
  \let\@ifdefinable\@@ifdefinable
  #2}
\def\newenvironment{\@star@or@long\new@environment}
\def\new@environment#1{%
  \@testopt{\@newenva#1}0}
\def\@newenva#1[#2]{%
   \@ifnextchar [{\@newenvb#1[#2]}{\@newenv{#1}{[#2]}}}
\def\@newenvb#1[#2][#3]{\@newenv{#1}{[#2][#3]}}
\def\renewenvironment{\@star@or@long\renew@environment}
\def\renew@environment#1{%
  \@ifundefined{#1}%
     {\@latex@error{Environment #1 undefined}\@ehc
     }{}%
  \expandafter\let\csname#1\endcsname\relax
  \expandafter\let\csname end#1\endcsname\relax
  \new@environment{#1}}
\long\def\@newenv#1#2#3#4{%
  \@ifundefined{#1}%
    {\expandafter\let\csname#1\expandafter\endcsname
                         \csname end#1\endcsname}%
    \relax
  \expandafter\new@command
     \csname #1\endcsname#2{#3}%
     \l@ngrel@x\expandafter\def\csname end#1\endcsname{#4}}

\def\providecommand{\@star@or@long\provide@command}
\def\provide@command#1{%
  {\escapechar\m@ne\xdef\@gtempa{{\string#1}}}%
  \expandafter\@ifundefined\@gtempa
    {\def\reserved@a{\new@command#1}}%
    {\def\reserved@a{\renew@command\reserved@a}}%
   \reserved@a}%

\def\@ifundefined#1{%
  \expandafter\ifx\csname#1\endcsname\relax
    \expandafter\@firstoftwo
  \else
    \expandafter\@secondoftwo
  \fi}

\chardef\@xxxii=32
\mathchardef\@Mi=10001
\mathchardef\@Mii=10002
\mathchardef\@Miii=10003
\mathchardef\@Miv=10004

\newcount\@tempcnta
\newcount\@tempcntb
\newif\if@tempswa\@tempswatrue
\newdimen\@tempdima
\newdimen\@tempdimb
\newdimen\@tempdimc
\newbox\@tempboxa
\newskip\@tempskipa
\newskip\@tempskipb
\newtoks\@temptokena

\long\def\@whilenum#1\do #2{\ifnum #1\relax #2\relax\@iwhilenum{#1\relax
     #2\relax}\fi}
\long\def\@iwhilenum#1{\ifnum #1\expandafter\@iwhilenum
         \else\expandafter\@gobble\fi{#1}}
\long\def\@whiledim#1\do #2{\ifdim #1\relax#2\@iwhiledim{#1\relax#2}\fi}
\long\def\@iwhiledim#1{\ifdim #1\expandafter\@iwhiledim
        \else\expandafter\@gobble\fi{#1}}
\long\def\@whilesw#1\fi#2{#1#2\@iwhilesw{#1#2}\fi\fi}
\long\def\@iwhilesw#1\fi{#1\expandafter\@iwhilesw
         \else\@gobbletwo\fi{#1}\fi}
\def\@nnil{\@nil}
\def\@empty{}
\def\@fornoop#1\@@#2#3{}
\long\def\@for#1:=#2\do#3{%
  \expandafter\def\expandafter\@fortmp\expandafter{#2}%
  \ifx\@fortmp\@empty \else
    \expandafter\@forloop#2,\@nil,\@nil\@@#1{#3}\fi}
\long\def\@forloop#1,#2,#3\@@#4#5{\def#4{#1}\ifx #4\@nnil \else
       #5\def#4{#2}\ifx #4\@nnil \else#5\@iforloop #3\@@#4{#5}\fi\fi}
\long\def\@iforloop#1,#2\@@#3#4{\def#3{#1}\ifx #3\@nnil
       \expandafter\@fornoop \else
      #4\relax\expandafter\@iforloop\fi#2\@@#3{#4}}
\def\@tfor#1:={\@tf@r#1 }
\long\def\@tf@r#1#2\do#3{\def\@fortmp{#2}\ifx\@fortmp\space\else
    \@tforloop#2\@nil\@nil\@@#1{#3}\fi}
\long\def\@tforloop#1#2\@@#3#4{\def#3{#1}\ifx #3\@nnil
       \expandafter\@fornoop \else
      #4\relax\expandafter\@tforloop\fi#2\@@#3{#4}}
\long\def\@break@tfor#1\@@#2#3{\fi\fi}
\def\@removeelement#1#2#3{%
  \def\reserved@a##1,#1,##2\reserved@a{##1,##2\reserved@b}%
  \def\reserved@b##1,\reserved@b##2\reserved@b{%
    \ifx,##1\@empty\else##1\fi}%
  \edef#3{%
    \expandafter\reserved@b\reserved@a,#2,\reserved@b,#1,\reserved@a}}

\let\ExecuteOptions\@gobble

\def\PackageError#1#2#3{%
  \errhelp{#3}\errmessage{#1: #2}}
\def\@latex@error#1#2{%
  \errhelp{#2}\errmessage{#1}}

\bgroup\uccode`\!`\%\uppercase{\egroup
\def\@percentchar{!}}

\ifx\@@input\@undefined
 \let\@@input\input
\fi

\def\input{\@ifnextchar\bgroup\@iinput\@@input}
\def\@iinput#1{\input{#1} }

    \def\filename@parse#1{%
      \let\filename@area\@empty
      \expandafter\filename@simple#1.\\}

  \def\filename@simple#1.#2\\{%
    \ifx\\#2\\%
       \let\filename@ext\relax
    \else
       \edef\filename@ext{\filename@dot#2\\}%
    \fi
    \edef\filename@base{#1}}
  \def\filename@dot#1.\\{#1}

\long\def \IfFileExists#1#2#3{%
  \openin\@inputcheck#1 %
  \ifeof\@inputcheck
    \ifx\input@path\@undefined
      \def\reserved@a{#3}%
    \else
      \def\reserved@a{\@iffileonpath{#1}{#2}{#3}}%
    \fi
  \else
    \closein\@inputcheck
    \edef\@filef@und{#1 }%
    \def\reserved@a{#2}%
  \fi
  \reserved@a}
\long\def\@iffileonpath#1{%
  \let\reserved@a\@secondoftwo
  \expandafter\@tfor\expandafter\reserved@b\expandafter
             :\expandafter=\input@path\do{%
    \openin\@inputcheck\reserved@b#1 %
    \ifeof\@inputcheck\else
      \edef\@filef@und{\reserved@b#1 }%
      \let\reserved@a\@firstoftwo%
      \closein\@inputcheck
      \@break@tfor
    \fi}%
  \reserved@a}
\long\def \InputIfFileExists#1#2{%
  \IfFileExists{#1}%
    {#2\@addtofilelist{#1}\@@input \@filef@und}}

\chardef\@inputcheck0

\let\@addtofilelist \@gobble

\def\@defaultunits{\afterassignment\remove@to@nnil}
\def\remove@to@nnil#1\@nnil{}

\newdimen\leftmarginv
\newdimen\leftmarginvi

\newdimen\@ovxx
\newdimen\@ovyy
\newdimen\@ovdx
\newdimen\@ovdy
\newdimen\@ovro
\newdimen\@ovri
\newdimen\@xdim
\newdimen\@ydim
\newdimen\@linelen
\newdimen\@dashdim

\long\def\mbox#1{\leavevmode\hbox{#1}}

\let\@onlypreamble\@gobble

\let\protect\relax

\newdimen\fboxsep
\newdimen\fboxrule

\def\@height{height} \def\@depth{depth} \def\@width{width}
\def\@minus{minus}
\def\@plus{plus}
\def\hb@xt@{\hbox to}

\long\def\@begin@tempboxa#1#2{%
   \begingroup
     \setbox\@tempboxa#1{\color@begingroup#2\color@endgroup}%
     \def\width{\wd\@tempboxa}%
     \def\height{\ht\@tempboxa}%
     \def\depth{\dp\@tempboxa}%
     \let\totalheight\@ovri
     \totalheight\height
     \advance\totalheight\depth}
\let\@end@tempboxa\endgroup

\let\set@color\relax
\let\color@begingroup\relax
\let\color@endgroup\relax
\let\color@setgroup\relax

\let\color@hbox\relax
\let\color@vbox\relax
\let\color@endbox\relax

\def\setlength#1#2{#1#2\relax}

\begingroup
  \catcode`P=12
  \catcode`T=12
  \lowercase{
    \def\x{\def\rem@pt##1.##2PT{##1\ifnum##2>\z@.##2\fi}}}
  \expandafter\endgroup\x
\def\strip@pt{\expandafter\rem@pt\the}

\def\@input#1{%
  \IfFileExists{#1}{\@@input\@filef@und}{\message{No file #1.}}}

\def\@warning{\immediate\write16}

\def\Gin@driver{dvips.def}
\input graphicx.sty

\resetatcatcode

\newcommand{\lissom}{\textsc{Lissom}\ }
\newcommand{\liss}{\textsc{Liss}\ }

\title{\textsc{Lissom}, a Source Level Proof Carrying Code Platform}%

\author{
  \begin{tabular}{cc}
    Jo\~{a}o Gomes \quad Daniel Martins \quad Sim\~{a}o Melo de
    Sousa\qquad \qquad  &  Jorge Sousa Pinto \\    
    Departamento de Inform\'{a}tica & DI/CCTC\\
    Universidade da Beira Interior &   Universidade do Minho\\ 
    Covilh\~{a}, Portugal & Braga, Portugal\\
    {\tt \{joao.gomes,daniel.martins,desousa\}@di.ubi.pt} & {\tt jsp@di.uminho.pt}
  \end{tabular}}

\date{}

\begin{document}

\maketitle 

\begin{abstract}
  
This paper introduces a proposal for a Proof Carrying Code (PCC)
architecture called \textsc{Lissom}.  Started as a challenge for final year
Computing students, \lissom was thought as a mean to prove to a
sceptic community, and in particular to students, that formal
verification tools can be put to practice in a realistic environment,
and be used to solve complex and concrete problems. The attractiveness
of the problems that PCC addresses has already brought students to
show interest in this project.

\end{abstract}

\section{The \lissom Platform}
\label{sec:LissomArchitecture}

Traditional PCC architectures center their certificate generation
mechanisms on the output of the compilation. Along the lines of
recent projects, we believe that there are strong benefits in moving
the certificate generation to the source code level. Because there
exist good tools for source code verification and for formal
verification in general, it is a feature of the \lissom platform that
existing tools are used as much as possible at key points of its
infrastructure.

Our vision of PCC is based on the following two underlying principles:
\begin{itemize}
\item \emph{Source level PCC is the way.}  It is our belief that the
  realistic formal verification of mobile code should be performed at
  source level. Programmers may be unaware of the target architecture
  details, and in general algorithmic constructions are expressed at
  source level.
\item \emph{Reuse as much as possible.}  There exist plenty of
  powerful tools for the formal verification of source code.  Such
  tools already have experienced user communities, and have reached an
  appreciable level of maturity and flexibility that make them natural
  choices in the context of a source level PCC architecture.
\end{itemize}
Our main goal with \lissom is to get experienced with PCC and to put
it into practice. We now describe the proposed architecture for the
\lissom platform (see figure below).

\paragraph{The Source Language and the  Compiler.}

LISS (Language for Integers Sets and Sequences) is a non-trivial toy
language that also features a realistic type system (with e.g. sets, 
vectors a la Java, etc.)
and high-level constructs. LISS is intended here as a suitable
test-bed before aiming at an industrial-level language. This language must be extended in order to provide an annotation system for the source code.  In a source
level PCC architecture, the compiler has to compile source code but
also proofs into their machine level counterpart. A very interesting
challenge is to transform a source level structural proof (proofs
heavily rely on the structure of the analyzed program) into a proof
that is still structurally close to the machine level code. We follow
here the contributions of~\cite{g+:certoc}). 

\begin{center}
\includegraphics[width=110mm]{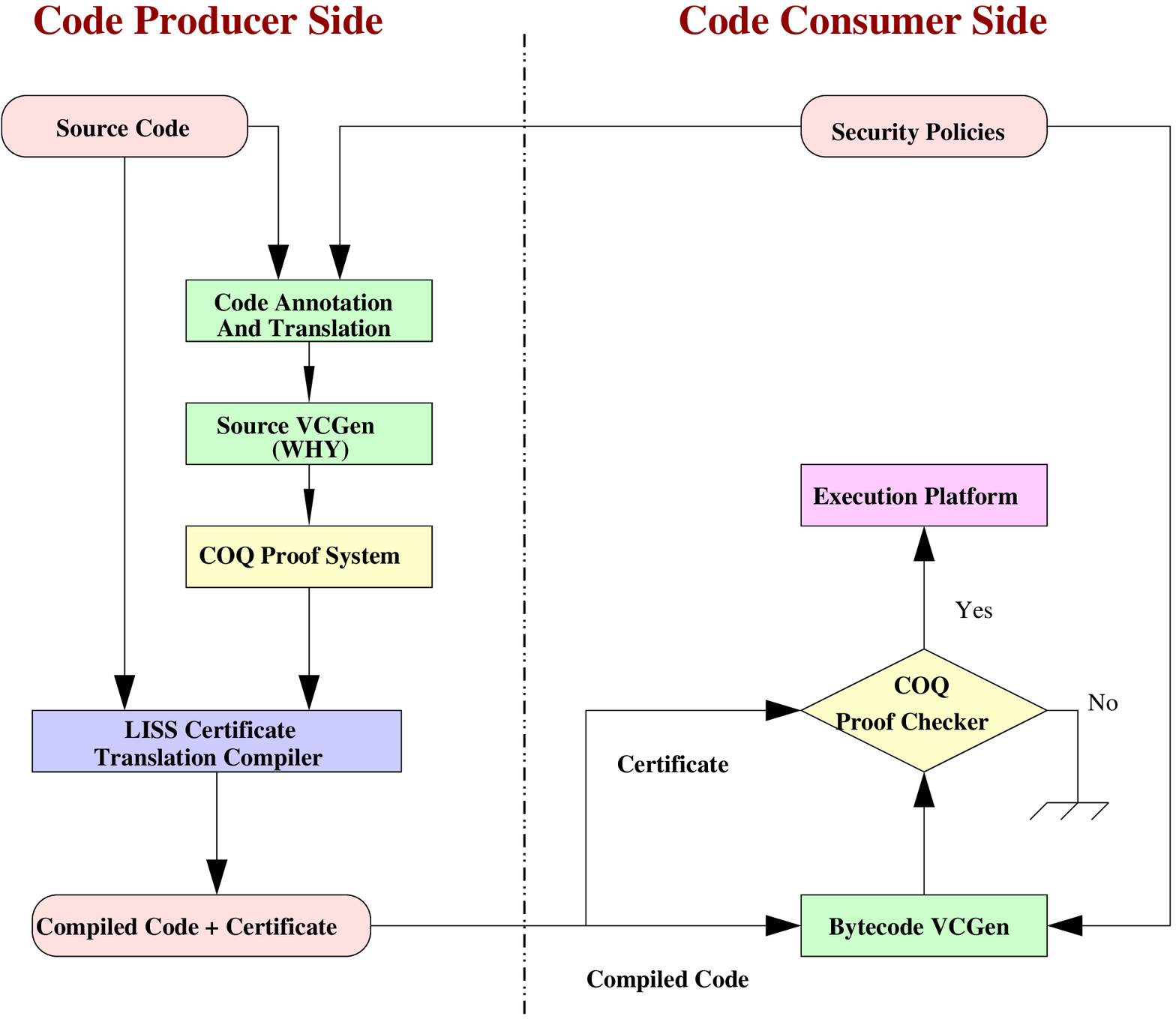}
\end{center}

\paragraph{The Virtual Machine.}

\lissom uses a sequential, stack-based virtual machine, which despite
its simplicity has the capacity to support real languages (such as C
or Java), and is, together with LISS, a suitable test-bed. The
machine is an adaptation of Filliatre's original virtual machine
\cite{VMurl}, used for teaching several courses at our universities
(e.g. Compiler Construction and Formal Methods).

\paragraph{The Proof System and the Proof Checker.}

As far as the \emph{Trusted Computing Base} (TCB) is concerned, it is
important for it to be as small and solid as possible; we believe that
an adequate choice of proof system may help attaining such a
goal. Also, it is important to be able to express high level
polices as well as lower level ones.

These requirements have led us to consider using the COQ proof system
and its higher-order specification language and underlying proof
mechanisms. This system, based on the calculus of inductive
constructions, has been used with success for the formal verification
of critical and large-scale systems. As far as source code is
concerned, integration with COQ is guaranteed by the existence of a
number of tools suited for code annotation and proof
(e.g.~\cite{FilliatreMarche04}). \lissom will feature a  source code
  verification system based on \textsc{why} and the \textsc{COQ}
  system. Thus we intend to use COQ proof objects as certificates.

\paragraph{The Verification Generator.}
\label{sec:TheVerificationGenerator}

This will be obtained using Filliatre's WHY
tool~\cite{Fill}, which is capable of producing proof
obligations for various systems, including COQ.
We are presently working on a WHY module for the language LISS
(equivalent to \emph{Caduceus} for C \cite{FilliatreMarche04}) and for
the input language of the virtual machine. The annotation language
used for this will be an adaptation of JML~\cite{BurdyEtAl04}
specialized for the security policy specification.

\section{Road Map}
\label{sec:Conlcusion}

After the present prototyping phase, the platform must have proved to
be adequate for mobile code security. The relevance and conceptual
solidity of previous works on formal verification (e.g.
\cite{g+:jakarta}) on which this is based lead us to believe in the
success of the enterprise. Our road-map is the following, where the points highlighted as \emph{in progress} are the modules which are currently in active development.

\begin{enumerate}
\item To design an annotation language for \liss (in progress);
\item to extend the \textsc{why} tool to contemplate this annotation
  language, allowing to use \textsc{why} as a generator of proof
  obligations for source code (in progress);
\item to design a proof system for the \liss language, integrated in
  \textsc{Coq} (starting phase);
\item to extend the \liss compiler with the capability to translate
  certificates (starting phase);
\item to design a proof system for the virtual machine and its
  language, integrated in \textsc{Coq} (in progress);
\item to design a proof obligation generator for compiled code (in progress);
\end{enumerate}
We finish with a few examples of the many interesting problems that
will be raised in this work, along with the classical challenges that every
PCC platform must address.  A first problem is the automation of the COQ
proof process and its impact on the consumer effort; the tension
between expressiveness and automation is a well-known problem that
must be carefully studied.
%
It also remains to see to what extent the techniques
presented in~\cite{NL98b} allow for conciseness of COQ certificates.

The language \liss and the virtual machine used are non-trivial, but
are still relatively simple when compared to platforms such as JAVA or
.NET. The capacity of the platform to scale up to such platforms must
thus be evaluated.
Finally, it will be important to apply our choices in an appropriate
case study that starts from the security policy specification to the
certificate verification (proof of concept).

\end{document}